\documentclass[a4paper,12pt]{article} 
\usepackage[T1]{fontenc}
\usepackage{times}
\usepackage{multicol}
\usepackage{hyperref}
\usepackage{amssymb}
\usepackage[sort&compress]{natbib}
\usepackage[switch]{lineno}
\usepackage{graphicx}
\usepackage{authblk}
\title{{\it Fermi}-LAT improved Pass~8 event selection}
\date{}
\author[1]{P.~Bruel}
\author[2]{T.~H.~Burnett}
\author[3]{S.~W.~Digel}
\author[4]{G.~J\'ohannesson}
\author[3]{N.~Omodei}
\author[3]{and M.~Wood}
\author[  ]{\\on behalf of the {\it Fermi}-LAT Collaboration}
\affil[1]{Laboratoire Leprince-Ringuet, \'Ecole polytechnique, CNRS/IN2P3, F-91128 Palaiseau, France}
\affil[2]{Department of Physics, University of Washington, Seattle, WA 98195-1560}
\affil[3]{W. W. Hansen Experimental Physics Laboratory, Kavli Institute for Particle Astrophysics and Cosmology, Department of Physics and SLAC National Accelerator Laboratory, Stanford University, Stanford, CA 94305, USA}
\affil[4]{Science Institute, University of Iceland, IS-107 Reykjavik, Iceland}

\begin{document}

\maketitle

\begin{abstract}
  The current version of the {\it Fermi} Large Area Telescope data (P8R2) has been publicly available since June 2015, with the caveat that the residual background of all event classes, except {\tt ULTRACLEANVETO}, was not fully isotropic: it was enhanced by a factor $\sim$2 at 1$-$3 GeV within $\sim$20~deg of the Ecliptic compared to the poles. By investigating the residual background using data only, we were able to find two sources of residual background: one due to non-interacting heavy ions and one due to cosmic-ray electrons leaking through the ribbons of the Anti-Coincidence Detector, the latter source being responsible for the background anisotropy. A set of simple cuts allows us to reject these events while losing less than 1\% of the {\tt SOURCE} class acceptance. This new selection has been used to produce a new version of the LAT data (P8R3).
\end{abstract}

\section{Introduction}
The {\it Fermi} Large Area Telescope (LAT)~\citep{latinstrument} has been collecting data since June 2008. As soon as the recorded events are downlinked, they are processed to estimate their energies, arrival directions, and many other quantities that are used in a multivariate analysis to assign them probabilities that they are photons. The event selection is based on these probabilities. The event reconstruction and selection have evolved since 2008 leading to 3 versions of the LAT data: first Pass~6 until August 2011, then Pass~7/Pass~7 reprocessed~\citep{pass7} until June 2015, when Pass~8~\citep{pass8} was released.

Contrary to Pass~7 for which only the selection had been changed, Pass~8 benefitted from a complete revision of the reconstruction allowing us to significantly improve the LAT performance: a larger acceptance, a more precise direction measurement as well as a better background rejection. Although the improvement from Pass~7 to Pass~8 was very clear, we noticed in Pass~8 a problem related to the spatial uniformity of the residual background that had not been detected in Pass~7 (we realized {\it a posteriori} that it was present in Pass~7, but at a much lower level than in Pass~8).

The LAT data are assigned to event classes corresponding to various event selections, each one ensuring a certain level of background. The {\tt SOURCE} event class is the most frequently used since it is the recommended event class for standard source analysis. To perform a standard LAT analysis of a region of the sky, one needs to define a model that comprises all sources that contribute to the photons in the region. It includes the various point and extended gamma-ray sources as well as the Galactic and isotropic diffuse emissions. The latter also includes the charged-particle residual background that is expected to be isotropic. Since the residual background changes with the event classes, we release a template of the isotropic diffuse emission for each event class~\footnote{\url{https://fermi.gsfc.nasa.gov/ssc/data/access/lat/BackgroundModels.html}}.

In Pass~8 the residual background of all event classes, except {\tt ULTRACLEANVETO}, was not isotropic: it was enhanced by a factor $\sim$2 at 1$-$3 GeV within $\sim$20~deg of the Ecliptic compared to the poles. The background nature of this anisotropy was demonstrated by the fact that it did not survive the {\tt ULTRACLEANVETO} very stringent selection. That is why the LAT collaboration recommended to use the {\tt SOURCE} event class for relatively small regions of interest ($\lesssim 25$~deg) and the {\tt ULTRACLEANVETO} event class for larger regions of the sky. The obvious drawback of this solution is that it implies a $\sim$25\% loss of acceptance for analyses of large regions of the sky. At the time of the Pass~8 release, whose version is referred to as P8R2, we did not investigate further the {\tt SOURCE} class anisotropy problem in order to avoid a delay of the release, but it was clear that mitigating it without losing acceptance would be of great value.

This proceedings paper~\footnote{This paper corresponds to a poster that was presented at the $8^\mathrm{th}$ International Fermi Symposium, October 14$-$19 2018, Baltimore.} describes the discovery of two sources of residual background in the {\tt SOURCE} event class (one being responsible for the background anisotropy) and the definition of a new event selection, leading to a spatially uniform {\tt SOURCE} residual background at the expense of a very small loss of acceptance. After presenting an improved photon/proton discriminating observable in Section~\ref{sec:sts}, we describe the first source of background in Section~\ref{sec:heavies}. Section~\ref{sec:method} presents a method to investigate the residual background based on data only ({\it i.e.}, without the help of the simulation) that led us to the discovery of the second source of background described in Section~\ref{sec:ribbons}. The new LAT event selection and its performance are presented in Section~\ref{sec:performance}.

\section{Shower transverse size} \label{sec:sts}

The LAT consists of three detector subsystems:
\begin{itemize}
\item a tracker/converter (TKR) in which the photon converts into an electron-positron pair whose direction is measured~\citep{TKR};
\item a hodoscopic CsI(Tl) crystal calorimeter (CAL) that measures the photon energy~\citep{CAL};
\item an Anti-Coincidence Detector (ACD) that tags incoming charged particles thanks to segmented tiles of scintillator~\citep{ACD} covering the $4\times4$ array of TKR+CAL towers.
\end{itemize}

Besides measuring the direction and the energy of the photons, the TKR and the CAL information is used to build observables that help to discriminate between photons and charged background particles such as electrons/positrons, protons and heavy ions. One of these observables is the logarithm of the transverse size of the shower in units of mm (STS) that allows us to disentangle electromagnetic showers (induced by either photons or electron/positrons) from hadronic showers. A precise measurement of the STS is made possible by the hodoscopic arrangement of the crystals in the CAL: each tower comprises 8 layers with 12 crystals per layer, with even and odd layer crystals having orthogonal orientations.

\begin{figure}[ht!]
  \centering
  \includegraphics[width=0.9\textwidth]{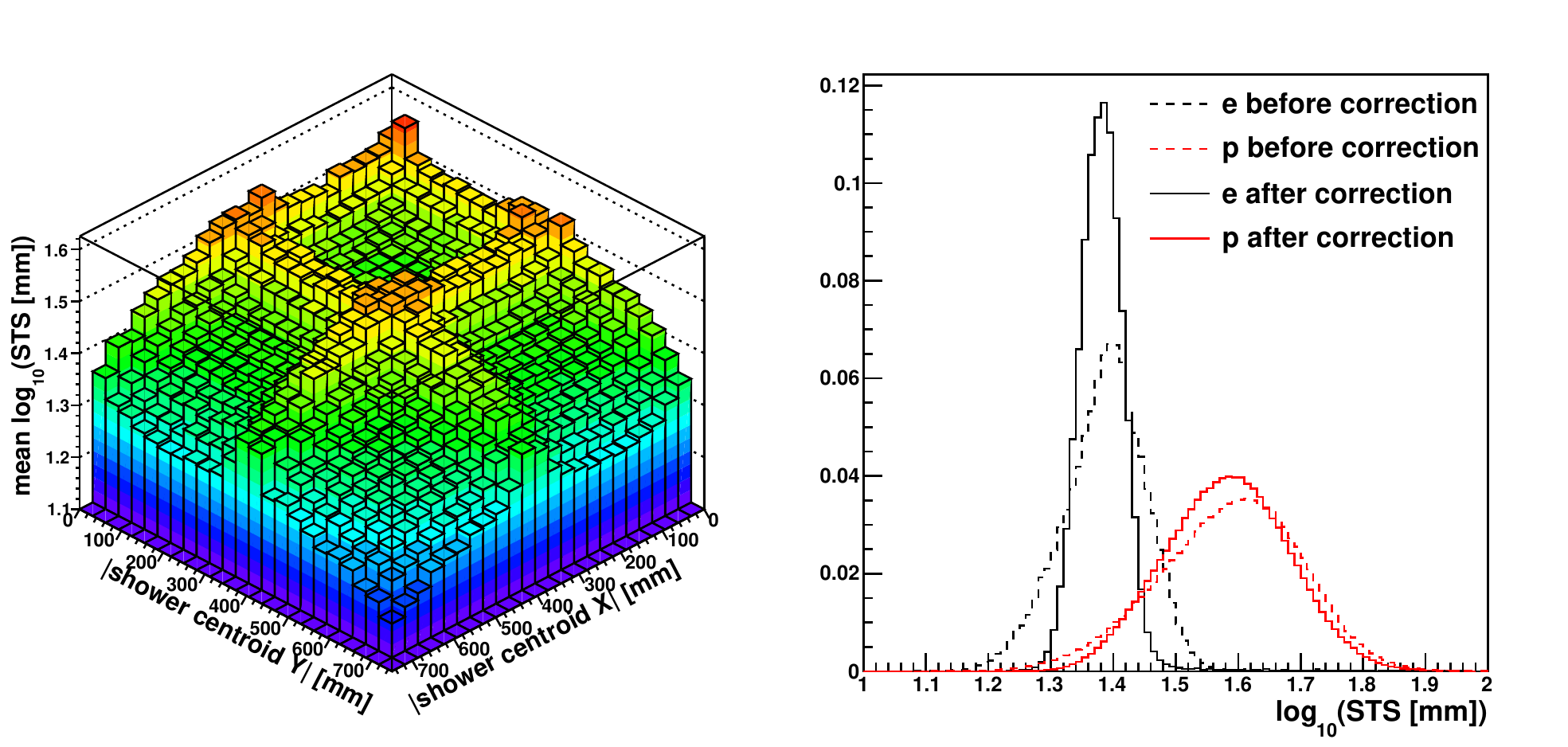}
  \caption{STS of 100<E<177~GeV electrons at all incidence angles: geometrical dependence (left) and electron/proton comparison before and after correction (right). For both electrons and protons, the datasets are the output of a simulation.}
  \label{fig:geomcor}
\end{figure}

In terms of photon selection, the STS is one of the key observables for proton rejection above $\sim$1~GeV. It becomes even more crucial for the electron selection since the ACD can not be used to discriminate between electrons and protons. That is why we tried to improve the STS discriminating power when deriving the electron selection for the measurement of the cosmic-ray electron spectrum~\citep{CRE}. A significant improvement was obtained by taking into account the geometrical dependence of the STS: the width and the containment of electron showers at a given energy vary with the incidence angle and entrance position in the CAL. In order to derive a simple correction for this geometrical dependence, we parameterize the geometrical dependence as a function of the absolute value of the $x$ and $y$ positions of the shower centroid with respect to the center of the LAT (we can use the absolute values because of the symmetry of the LAT), based on an electron simulation performed with the Geant4 package~\citep{GEANT4}. An example of such dependence is shown in Figure~\ref{fig:geomcor}~(left) for electrons between 100 and 177~GeV, in which the size of the $(|x|,|y|)$ plane corresponds to one quarter of the LAT, {\it i.e.,} $2\times2$ towers. The STS variations clearly follow the tower structure of the LAT, with the STS being on average larger when the shower centroid is close to the gaps between towers. Since the geometrical dependence varies with energy, the correction is also parametrized as a function of energy.

Figure~\ref{fig:geomcor}~(right) compares the STS distribution of electrons and protons whose reconstructed energy is between 100 and 177~GeV before and after the correction.  One can see that the STS distribution is much narrower after correction for electrons, which improves significantly its discriminating power: for a 90\% cut efficiency, the background rejection is increased by a factor larger than 2.5.

To take advantage of the STS correction in the context of the photon/proton discrimination, the same geometrical correction is derived for photons, except that we produce two sets of corrections corresponding to events converting in the front (top) and back (bottom) parts of the tracker. In the rest of the paper, STS refers to the logarithm of the shower transverse size after correction for its geometrical dependence.

\section{Non-interacting heavy ions} \label{sec:heavies}

Another lesson from the cosmic-ray electron spectrum analysis was that above $\sim$10~GeV, the average time-over-threshold in the tracker (ToT, in units of minimum ionizing particle) helps in distinguishing electromagnetic showers from hadronic showers, because it provides a measurement of the energy deposited at the beginning of the shower. Hadronic showers tend to develop later and less rapidly than electromagnetic showers. So their STS is artificially reduced because, at a given reconstructed energy, a smaller fraction of the shower is contained in the CAL compared to electromagnetic showers. The ToT can be used to select events for which the shower starts in the tracker, increasing on average the proton STS while the photon/electron STS does not change much.

\begin{figure}[ht!]
  \centering
  \includegraphics[width=0.9\textwidth]{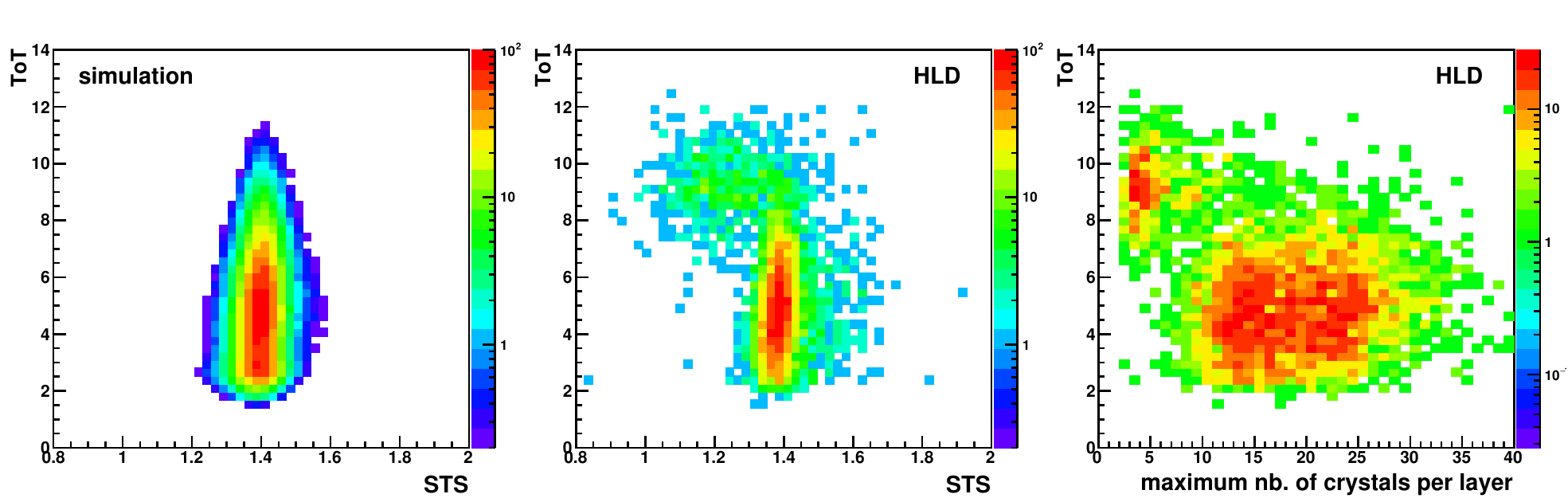}
  \caption{The ToT vs. STS distributions of 80<E<125~GeV {\tt SOURCE} events: simulation (left) and HLD events (center). The ToT vs. the maximum number of crystals per layer for the same HLD events (right).}
  \label{fig:heavies}
\end{figure}

The region of the gamma-ray sky with the largest relative background contamination corresponds to the high Galactic latitudes, excluding the {\it Fermi} bubbles~\citep{bubblesDobler,bubblesSu,bubblesLAT} and the known gamma-ray sources. We consider a background-rich data sample defined as high Galactic latitude events ($|b|>20$~deg and excluding the {\it Fermi} bubbles) that are more than 0.3~deg away from 3FGL sources, and refer to it as the high latitude Galactic diffuse emission (HLD). When looking at the ToT vs STS distribution for HLD {\tt SOURCE} events, we serendipitously discovered an accumulation of events with a large ToT and a very small STS, which is not expected for photons. Figure~\ref{fig:heavies} shows this distribution for events between 80 and 125~GeV for data and photon simulation. The cluster of events at small STS with 7<ToT<12 seen with the HLD events is not present in the simulation.

Figure~\ref{fig:heavies} also shows the distribution of ToT vs. the maximum number of hit crystals ({\it i.e.}, with more than a few MeV deposited energy) per layer of the high Galactic latitude events. One can see that the large ToT events have a maximum number of crystals per layer much smaller than the bulk of the events. Another feature of these events is that the energies deposited in the 8 layers of the CAL are very similar, contrary to what is expected for electromagnetic showers developing in the CAL. Because of their small maximum number of crystals per layer and the almost constant energy loss rate through the layers (that is larger than the one expected for minimum ionizing protons), these events are very likely due to minimum ionizing heavy ions.

We apply a simple cut based on the ToT and the maximum number of the calorimeter crystals per layer to reject these heavy ion induced events. Most of them have a reconstructed energy between 30 and 300~GeV. The fraction of these events out of the high latitude Galactic diffuse emission peaks around 100~GeV, reaching almost 15\%. This cut is applied in the rest of the paper. We note that, after this cut is applied, we do not expect any significant background in the leading edge of the STS distribution.

\section{Residual background investigation} \label{sec:method}

The STS is a very useful observable to discriminate between electromagnetic showers and hadronic showers but, as shown in the cosmic-ray electron spectrum analysis~\citep{CRE}, it is not perfectly reproduced by the LAT simulation. As a result it is not straightforward to derive very stringent STS-based background rejection cuts directly from the simulation. However, the STS can be used to study the residual background using data only.

For that purpose, one needs to compare two {\tt SOURCE} class data samples: one that is very rich in gamma-ray photons and that can serve as reference, and one that contains more background compared to the reference dataset. In this analysis, the reference sample (REF) comprises three subsamples: events from the Earth limb~\footnote{Earth limb gamma-ray emission is due to cosmic-ray protons interacting at the top of the atmosphere at near grazing incidence.}, from the Galactic plane ($|b|<5$~deg and $|l|<90$~deg) and those within 0.2~deg of a 3FGL source. The background-rich sample corresponds to the high latitude Galactic diffuse emission (HLD) defined in the previous Section.

In order to investigate the residual background using data only, we compare the distributions of photon/proton discriminating observables between the two event samples. We use two observables: the STS and $S_\mathrm{tile}$, the logarithm of the signal in the ACD tile (in units of minimum ionizing particle) to which the event direction points.
      
We define a ``pure-photon'' region (PPR) in the (STS,$S_\mathrm{tile}$) plane by requiring a very small shower transverse size and no ACD tile signal. We assume that the background is negligible in the PPR and we use the PPR to renormalize the REF sample so that the REF and HLD samples have the same number of events in the PPR. After renormalization, the total difference of events between the HLD and REF samples gives the number of residual background events.

Since the HLD sample contains more background than the REF sample, all HLD distributions are expected to have more events than the REF ones in every part of the phase-space. The contrary would mean that the normalization based on the PPR is incorrect, which would imply that the PPR contains a significant amount of background and that the definition of the PPR must be modified.

\begin{figure}[ht!]
  \centering
  \includegraphics[width=0.9\textwidth]{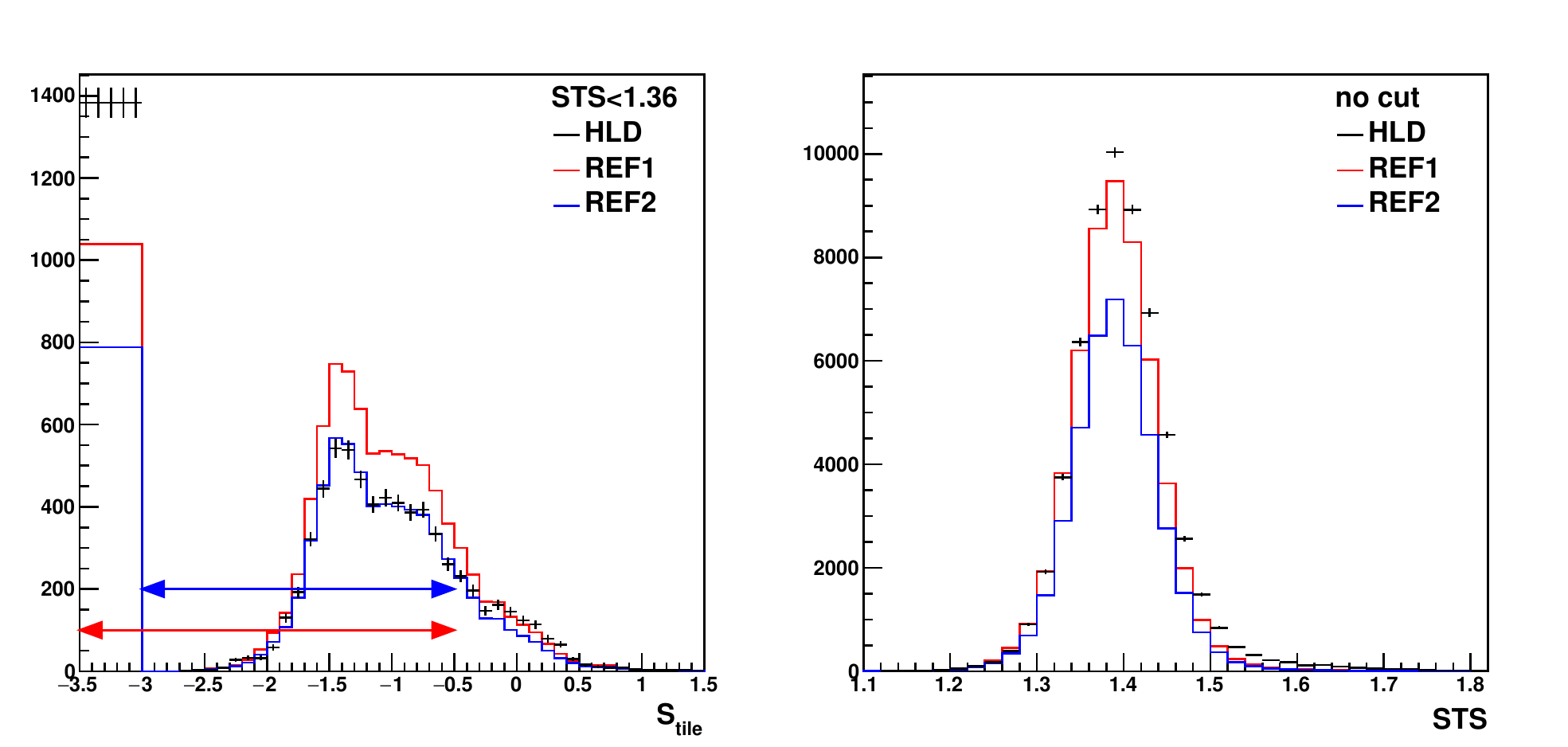}
  \caption{Events with 10<E<13~GeV: $S_\mathrm{tile}$ distribution for STS<1.36 (left) and STS distribution without cut (right). The black points correspond to the HLD sample, while the red and blue histograms correspond to the REF sample for two definitions of the PPR, indicated by the red and blue arrows. In the $S_\mathrm{tile}$ distribution, events with no signal have been evenly distributed among the first five bins of the histograms to better show the distribution for $S_\mathrm{tile}>-3$.}
  \label{fig:method}
\end{figure}

We start defining the PPR by requiring that the events have a very small shower transverse size (STS<1.36) and that the ACD tile signal is well below what is expected for a charged particle ($S_\mathrm{tile}<-0.5$). Figure~\ref{fig:method}~(left) shows the $S_\mathrm{tile}$ distribution of the HLD (black) and REF (red/REF1) sample for 10<E<13~GeV events with STS<1.36. One can see that there are more REF events than HLD events in the range $[-1.8,-0.5]$, which should not occur.

Changing the PPR definition to not include the events with no ACD tile signal at all ({\it i.e.}, events with $S_\mathrm{tile}<-3$) solves the problem: the new REF distribution (blue/REF2) matches the HLD distribution very well in the PPR range.

Figure~\ref{fig:method}~(right) shows the STS distribution of all the 10<E<13~GeV events. The difference between the HLD and REF histograms corresponds to the residual background contained in the HLD sample. The first but incorrect PPR definition (red) gives an estimate of the fraction of the residual background in the HLD of about 10\%. The correct fraction is $\sim$30\%, as obtained with the second PPR definition (blue).

\section{Cosmic-ray electrons/positrons through ribbons} \label{sec:ribbons}

\begin{figure}[ht!]
  \centering
  \includegraphics[width=0.9\textwidth]{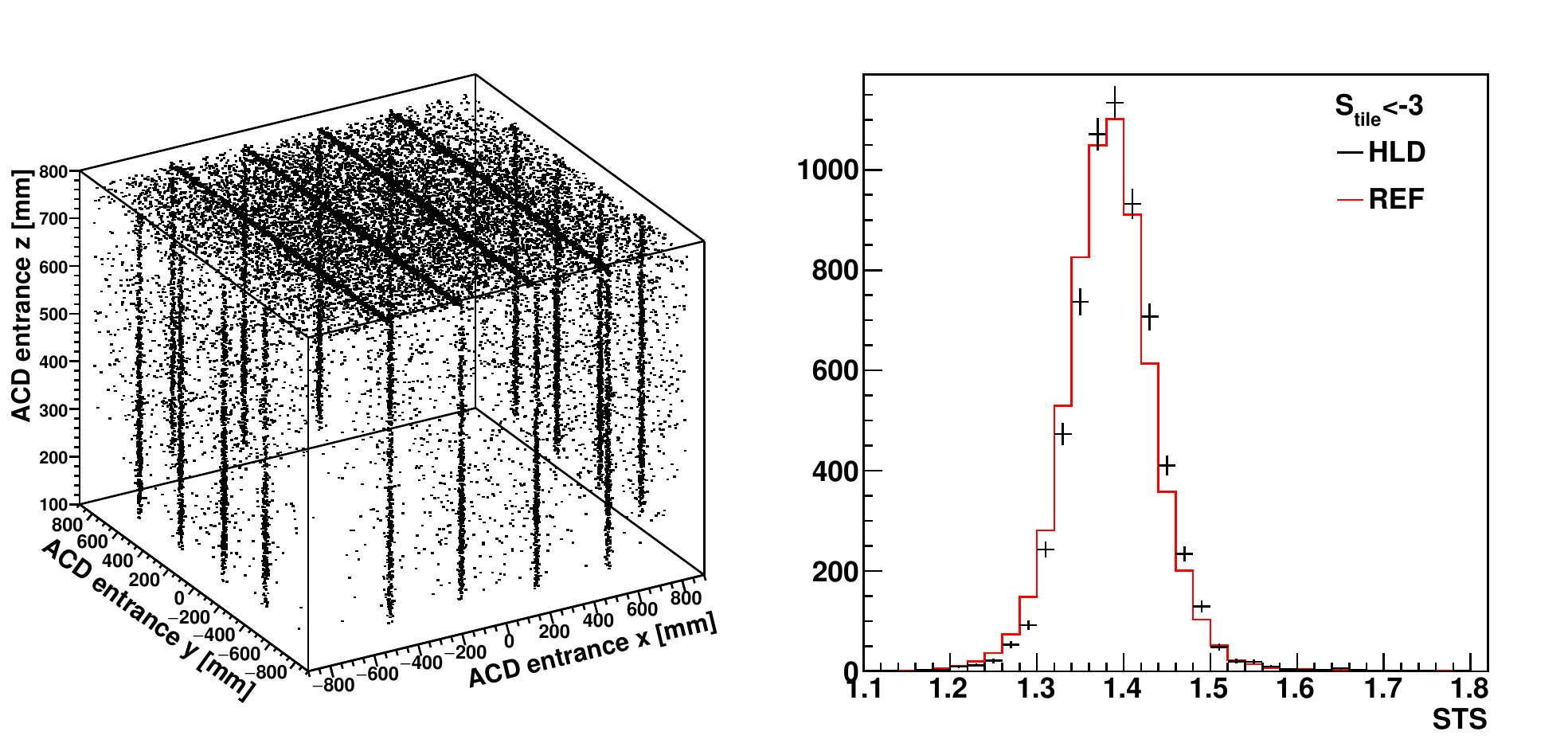}
  \caption{HLD 10<E<13~GeV events without any ACD tile signal: distribution of the $(x,y,z)$ coordinates of the ACD entrance point (left) and STS distribution compared to the STS distribution of REF events (right).}
  \label{fig:ribbon}
\end{figure}

The distribution of the ACD entrance point of HLD events without any ACD tile signal, as shown for 10<E<13~GeV events in Figure~\ref{fig:ribbon}~(left), demonstrates that most of them pass through the scintillating fiber ribbons of the ACD. To ensure a hermetic ACD, the tiles are overlapped in one dimension and the gaps along the other direction are covered by eight flexible scintillating fiber ribbons. They are $\sim$3~m long and $\sim$1.3~cm wide and read out at both ends by photo-multipliers. As a matter of fact, a signal in one of the ribbons is measured for most of these events, which proves that they are due to charged particles. Figure~\ref{fig:ribbon}~(right) compares the STS distribution of these events to the distribution of the REF events that also have no ACD tile signal. One can see that these two distributions are very similar, which proves that these background events produce an electromagnetic shower in the CAL. Since they are charged, they are cosmic-ray electrons or positrons. In the rest of the paper we refer to them as electrons, for simplicity's sake.

One could wonder why these events passed the {\tt SOURCE} class selection in the first place. The {\tt SOURCE} class selection, as well as the other LAT event classes, is based on a multivariate analysis performed with TMVA~\citep{TMVA}. This is done by training Boosted Decision Trees with simulated datasets on a set of observables that have some photon/background discriminating power. The signals in the ACD tile and ribbon are both among that set of observables but most of background events have a tile signal and no ribbon signal and, moreover, the ribbon information is not as easy to use as that from the tiles (because of the ribbon geometry and the signal attenuation along them). So it is very likely that the multivariate analysis gives much more weight to the tile information than to the information from the ribbons and therefore fails to reject efficiently electrons passing through the ribbons since, as soon as they passed the ACD, there is no way to disentangle them from photons.

For an electron to pass through a ribbon without depositing any energy in a tile, its incoming direction must lie in the plane perpendicular to the LAT side it went through and containing the ribbon. As a consequence, although the cosmic-ray electron flux is isotropic, the azimuthal distribution in the instrument frame of these ribbon-leaked electrons presents four peaks (at 0, 90, 180 and 270~degrees). Because the LAT is oriented such that the +X side faces the Sun most of the time, a large part of this electron background is reconstructed as originating from a direction near the Ecliptic.

We have derived an energy-dependent cut on the ACD ribbon signal so that the amplitude of the residual four-peak structure of the azimuthal distribution due to ribbon leakage is less than 1\% of the HLD flux.

\section{New version of the Pass~8 event selection} \label{sec:performance}

The cuts that remove the minimum ionizing heavy ions and the ribbon-leaked electrons have been added to the P8R2 event selection for all existing event classes in P8R2. All the LAT data have been reclassified with the resulting new Pass~8 event selection (P8R3).

\begin{figure}[ht!]
  \centering
  \includegraphics[width=0.9\textwidth]{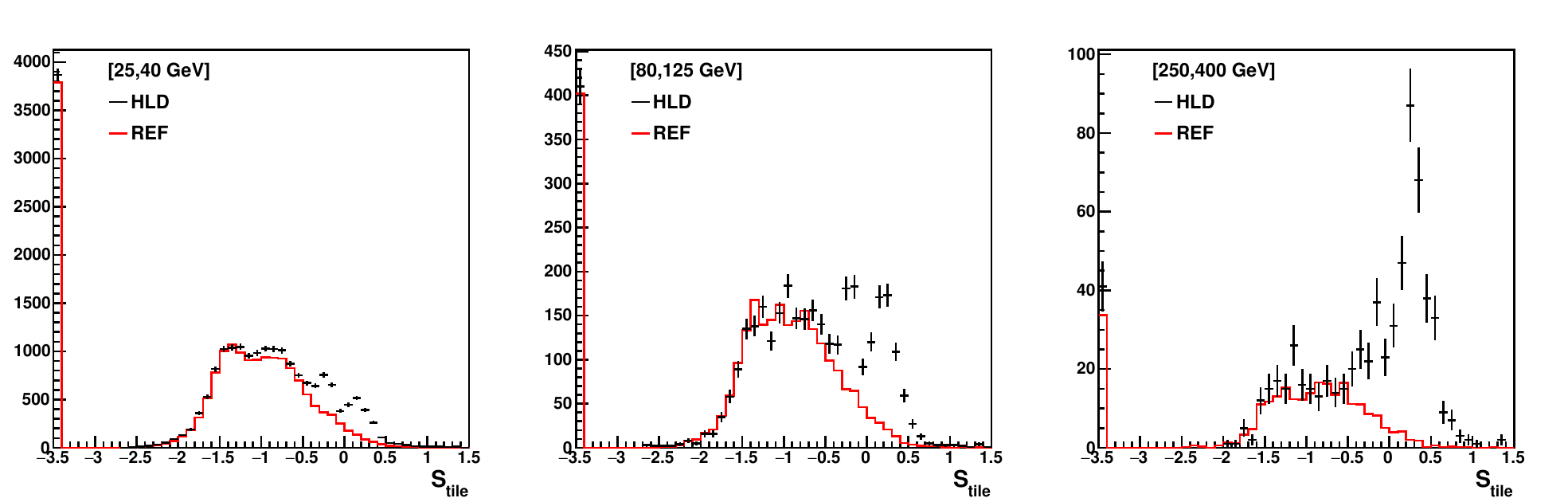}
  \caption{$S_\mathrm{tile}$ distribution for events without any ribbon signal.}
  \label{fig:sourceveto}
\end{figure}

The HLD/REF comparison method presented in Section~\ref{sec:method} has been used to investigate the residual background in the P8R3 {\tt SOURCE} class. The results are shown in Figure~\ref{fig:sourceveto} for three energy bins, where it can be seen that the residual background events have $S_\mathrm{tile}>-0.5$ and seem to fall into two categories of events (with $S_\mathrm{tile}$ less than and greater than 0).

After investigation it was found that the two categories actually correspond to whether there is one non-hit TKR silicon strip detector (SSD) between the head of the track and the ACD. Events with at least one non-hit TKR SSD have $S_\mathrm{tile}<0$ and events without have $S_\mathrm{tile}>0$. This separation into two categories is in fact not surprising: from the point of view of the multivariate analysis, events with at least one non-hit TKR SSD between the head of the track and the ACD more likely correspond to neutral particles and a stringent cut on the ACD tile signal is not essential. To the contrary, events without any non-hit TKR SSD between the head of the track and the ACD more likely correspond to charged particles and a stringent cut on the ACD tile signal is then needed.

As a consequence, we have implemented a new event class named {\tt SOURCEVETO}, that corresponds to the {\tt SOURCE} selection with a cut on the ACD tile signal that depends on whether or not there is one non-hit TKR SSD between the head of the track and the ACD. The purpose of this new event class is to reach a low level of residual background above 10~GeV while having a larger acceptance than the {\tt ULTRACLEANVETO} class.

\begin{figure}[ht!]
  \centering
  \includegraphics[width=0.8\textwidth]{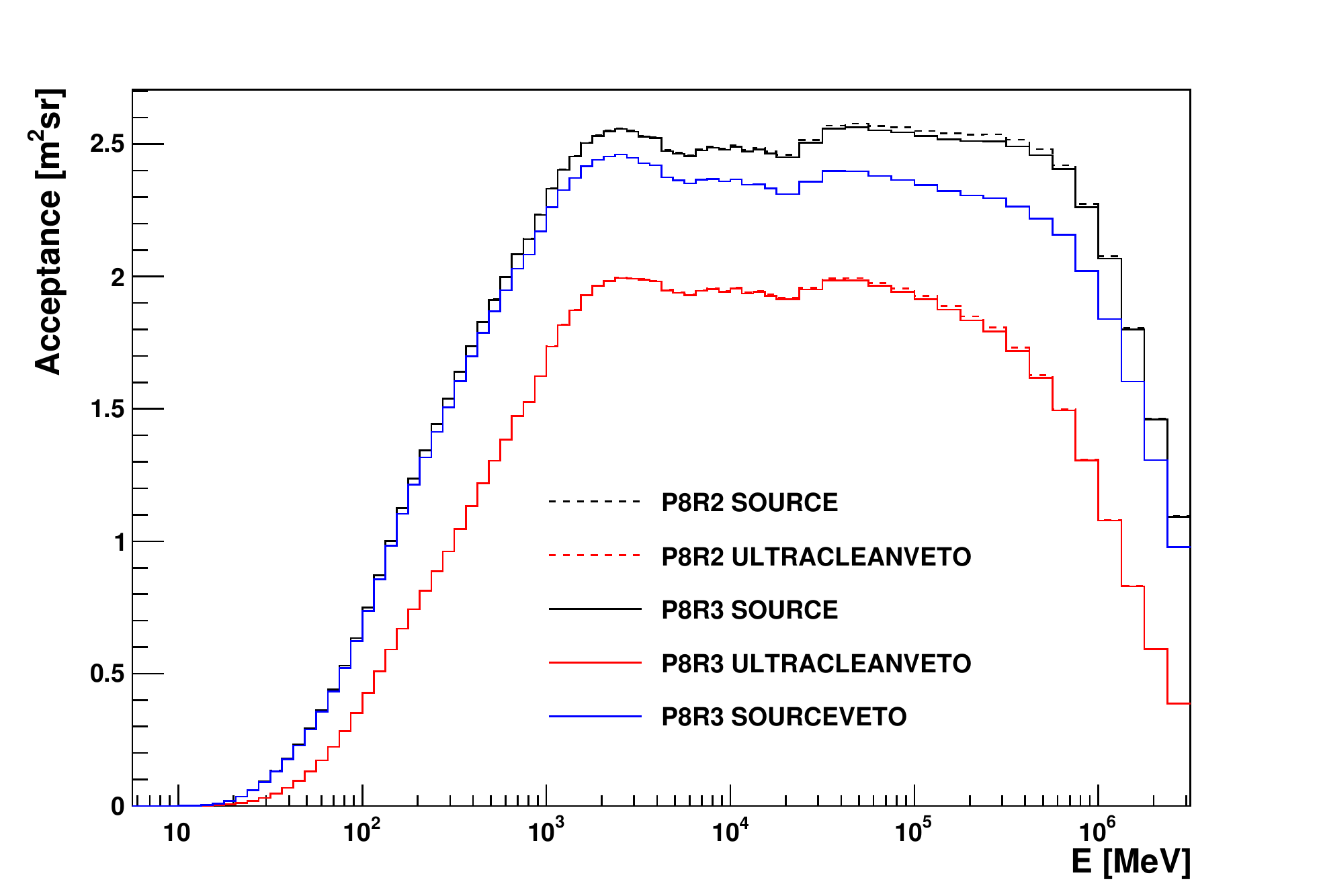}
  \caption{Comparison of acceptances for some P8R2 and P8R3 event classes.}
  \label{fig:acceptance}
\end{figure}

Figure~\ref{fig:acceptance} shows the acceptance of some P8R2 and P8R3 event classes. The loss of acceptance of the {\tt SOURCE} class between P8R2 and P8R3 is very small: it is negligible below 10~GeV and reaches barely 1\% at $\sim$~300~GeV. The loss of acceptance for the {\tt ULTRACLEANVETO} selection is even smaller. The acceptance of the new {\tt SOURCEVETO} class is about 10\% less than for {\tt SOURCE}.

\begin{figure}[ht!]
  \centering
  \includegraphics[width=0.8\textwidth]{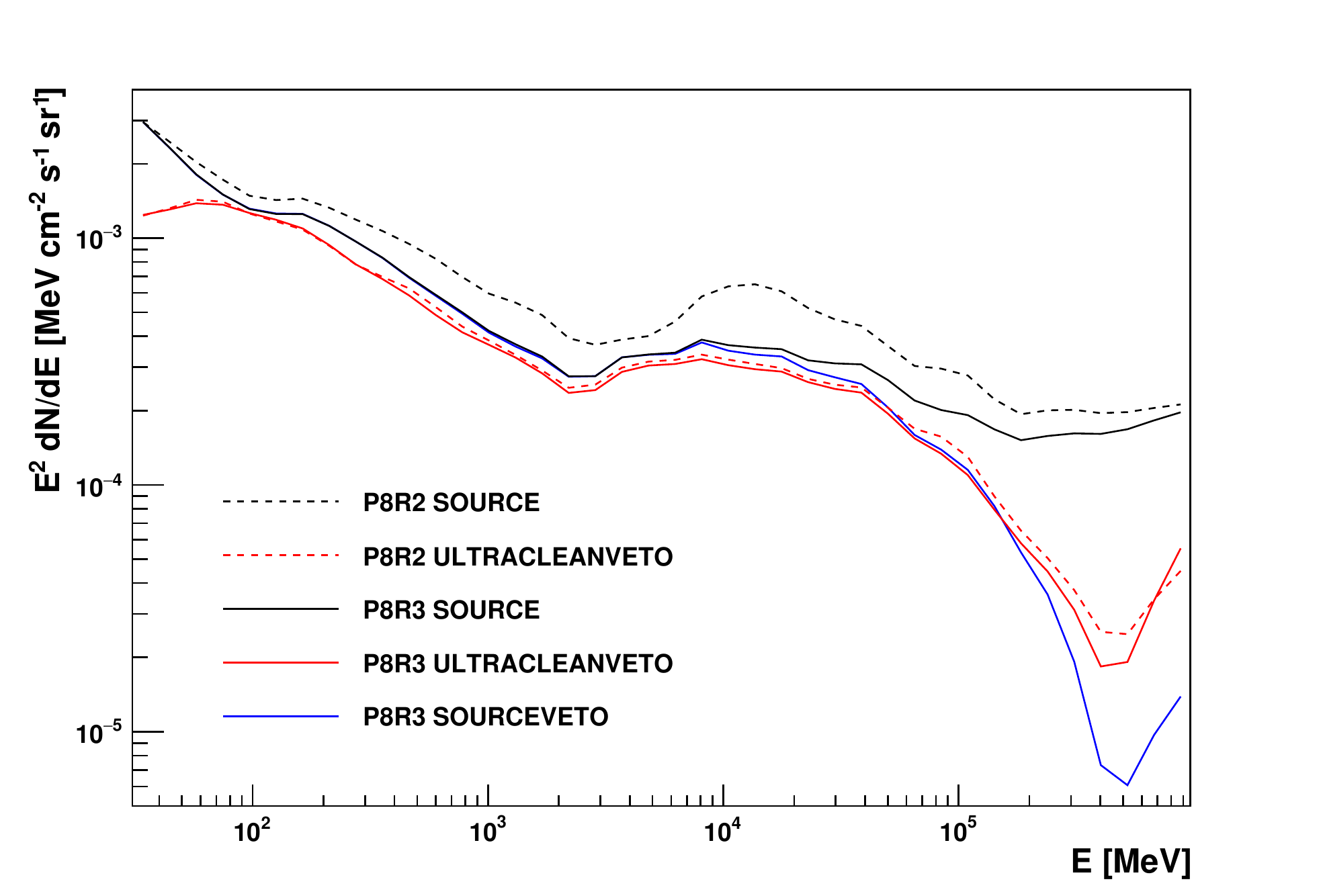}
  \caption{Isotropic template comparison for some P8R2 and P8R3 event classes.}
  \label{fig:isotemplates}
\end{figure}

Figure~\ref{fig:isotemplates} compares the {\tt SOURCE} and {\tt ULTRACLEANVETO} isotropic templates that were derived assuming the current LAT Galactic diffuse emission model~\footnote{\url{https://fermi.gsfc.nasa.gov/ssc/data/access/lat/BackgroundModels.html}}~\citep{GALDIF}. The {\tt SOURCE} residual background has decreased very much from P8R2 to P8R3 and is very close to the {\tt ULTRACLEANVETO} residual background below $\sim$20~GeV. One can see that above $\sim$50~GeV, the residual background of the new {\tt SOURCEVETO} class is equal to or less than the {\tt ULTRACLEANVETO} class, although it has a larger acceptance.

\begin{figure}[ht!]
  \centering
  \includegraphics[width=0.4\textwidth]{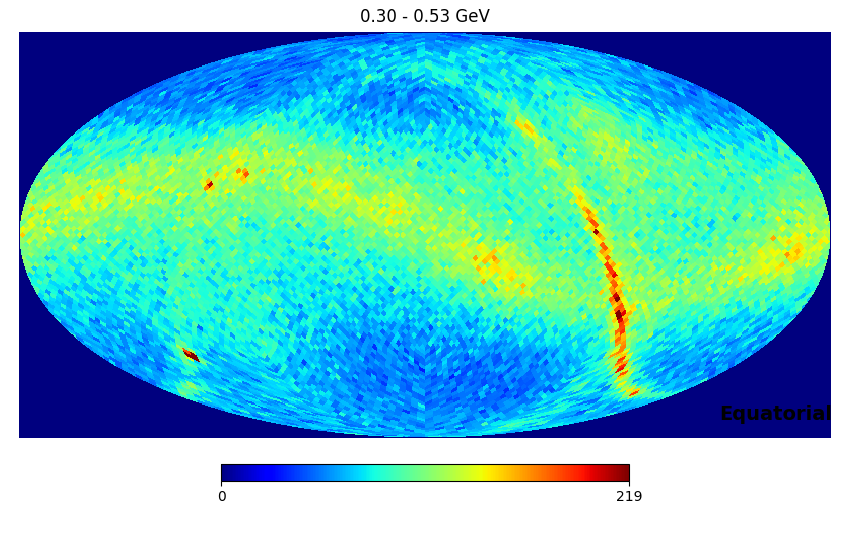}~\includegraphics[width=0.4\textwidth]{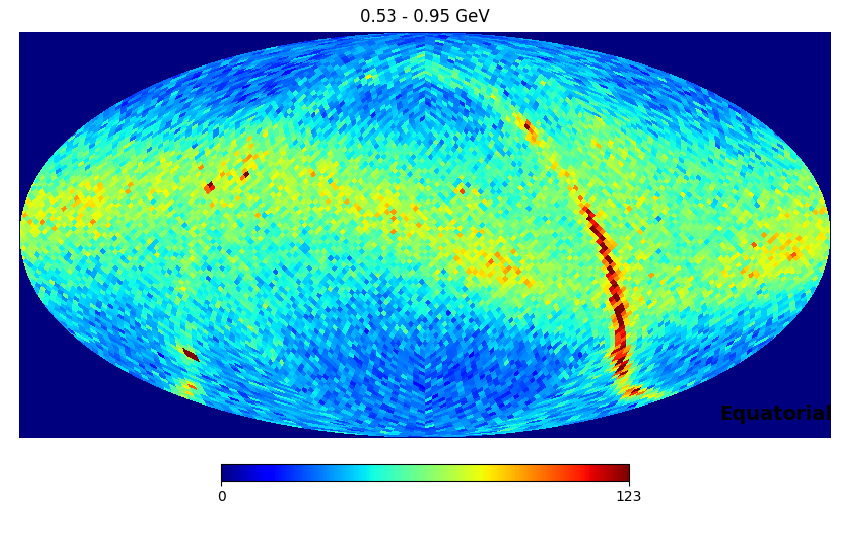}
  \includegraphics[width=0.4\textwidth]{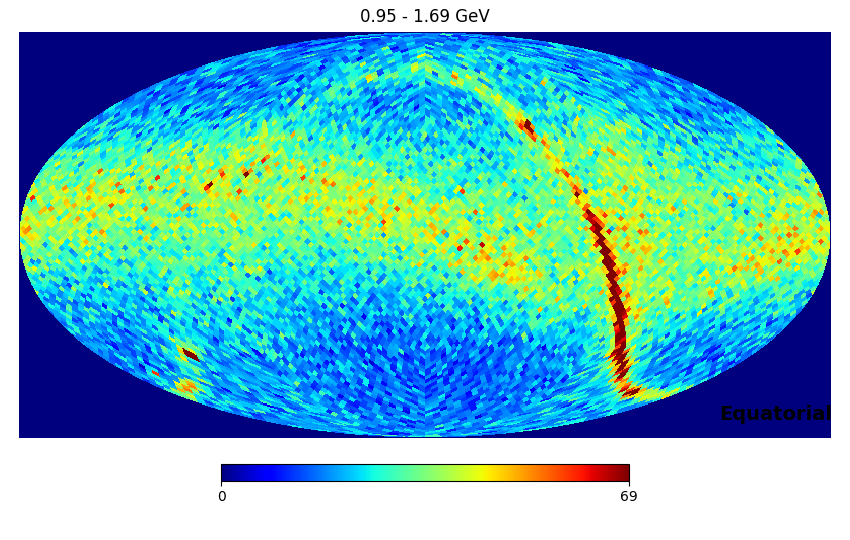}~\includegraphics[width=0.4\textwidth]{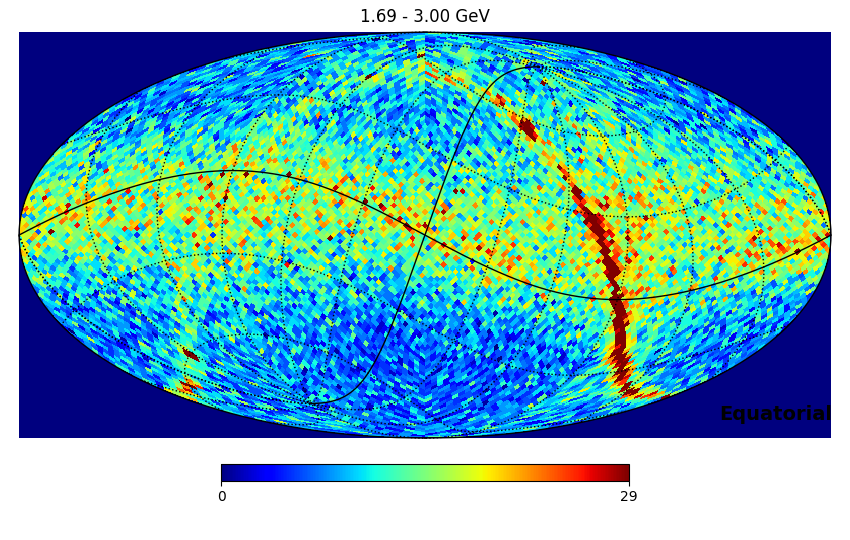}
  \caption{Maps in equatorial coordinates for the energy ranges indicated of the P8R2 {\tt SOURCE} events that are rejected by the P8R3 {\tt SOURCE} selection. Overlaid on the 1.69$-$3.00 GeV plot is an ecliptic coordinate grid.  In each image the brightest band is gamma rays in the Galactic plane, owing to the slightly smaller acceptance for gamma rays of P8R3. In addition to the bright diffuse band around the ecliptic, the P8R2-P8R3 images show a possible fainter component perpendicular to the equatorial and ecliptic planes.}
  \label{fig:p302-p305}
\end{figure}

\begin{figure}[ht!]
  \centering
  \includegraphics[width=0.8\textwidth]{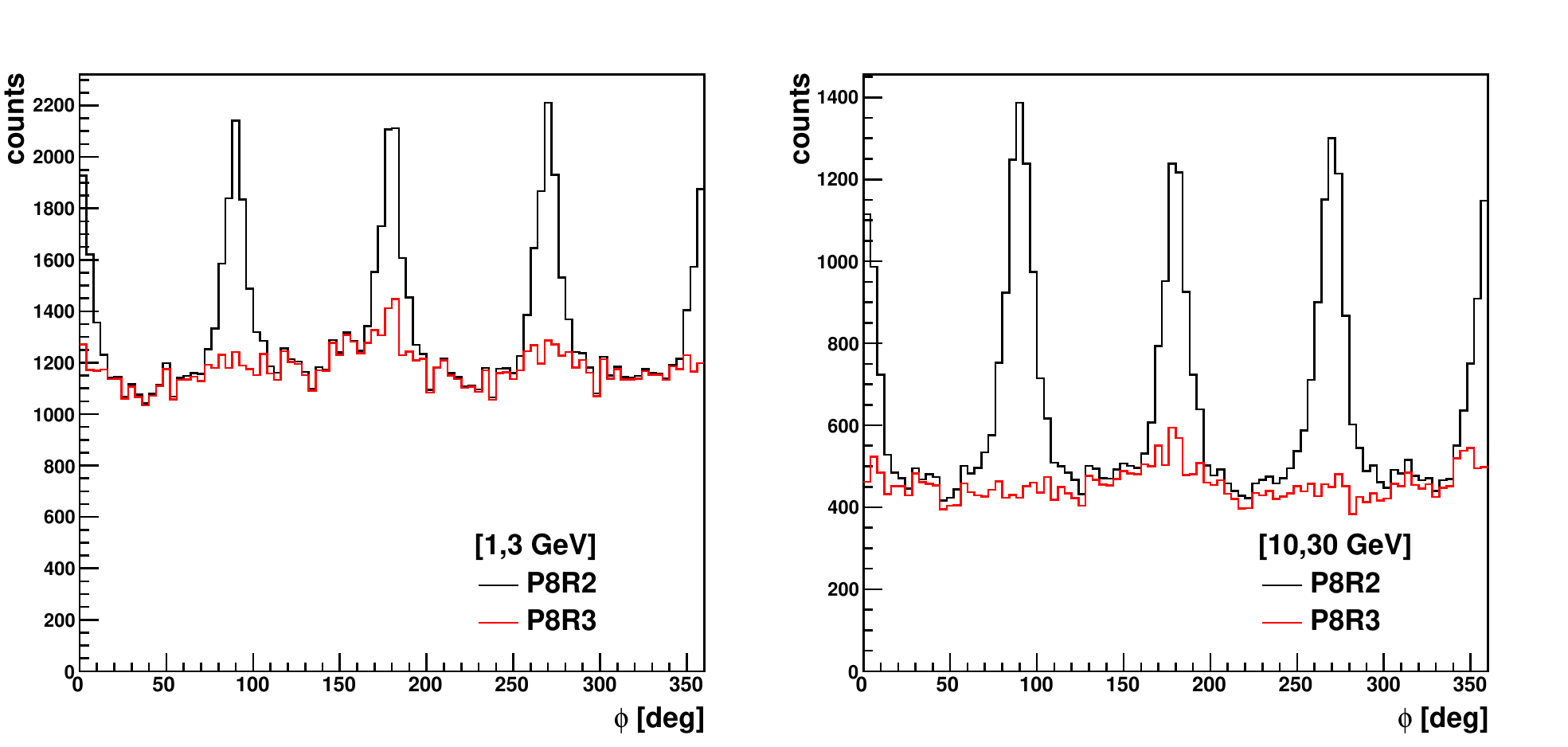}
  \caption{Azimuthal distribution in the instrument frame of very high Galactic latitude SOURCE events for P8R2 (black) and P8R3 (red) in two energy bins. The time ranges are 1.7 and 9 years for the [1,3~GeV] and the [10,30~GeV] energy bins, respectively.}
  \label{fig:phi}
\end{figure}

The difference of {\tt SOURCE} events between P8R2 and P8R3 is about 4\% below 10~GeV and about 10\% above 10~GeV. Maps of the P8R2 {\tt SOURCE} events rejected by the P8R3 event selection are shown in Figure~\ref{fig:p302-p305} for four energy bins. The Ecliptic-like structure is very clear. Figure~\ref{fig:phi} compares the P8R2 and P8R3 azimuthal distributions of SOURCE events at very high Galactic latitude ($|b|>50$~deg) that are not within three times the point spread function 68\% containment of any 3FGL source. The four-peak structure is greatly suppressed in P8R3, with a residual amplitude consistent with the azimuthal dependence of the photon effective area.


\section{Conclusion}

We have presented an investigation of residual charged particle background in the Pass~8 {\tt SOURCE} event class that led to a new event selection (P8R3) with overall less background and especially no significant anisotropic background. These new LAT data have already been used in the derivation of the preliminary 8-year {\it Fermi}-LAT source list (FL8Y)~\footnote{\url{https://fermi.gsfc.nasa.gov/ssc/data/access/lat/fl8y}} and are being used to prepare the next {\it Fermi}-LAT catalog of gamma-ray sources (4FGL).

\section*{Acknowledgments}

The \textit{Fermi} LAT Collaboration acknowledges generous ongoing support
from a number of agencies and institutes that have supported both the
development and the operation of the LAT as well as scientific data analysis.
These include the National Aeronautics and Space Administration and the
Department of Energy in the United States, the Commissariat \`a l'Energie Atomique
and the Centre National de la Recherche Scientifique / Institut National de Physique
Nucl\'eaire et de Physique des Particules in France, the Agenzia Spaziale Italiana
and the Istituto Nazionale di Fisica Nucleare in Italy, the Ministry of Education,
Culture, Sports, Science and Technology (MEXT), High Energy Accelerator Research
Organization (KEK) and Japan Aerospace Exploration Agency (JAXA) in Japan, and
the K.~A.~Wallenberg Foundation, the Swedish Research Council and the
Swedish National Space Board in Sweden.
 
Additional support for science analysis during the operations phase is gratefully
acknowledged from the Istituto Nazionale di Astrofisica in Italy and the Centre
National d'\'Etudes Spatiales in France. This work performed in part under DOE
Contract DE-AC02-76SF00515.

\end{document}